\documentstyle[eqsecnum,aps,prb]{revtex}

\def\s{\sigma}
\def\l{\ell}

\draft
\preprint{}
\begin{document}

\title{Mechanism of resonant x-ray magnetic scattering in NiO}

\author{Jun-ichi Igarashi and Manabu Takahashi}

\address{Faculty of Engineering, Gunma University, Kiryu, 
Gunma 376-8515, Japan}

\date{ \ \ \ \ \ \ \ \ \ \ \ \ \ \ \ \ \ \ \
}
\maketitle

\def\vol(#1,#2,#3){{\bf #1}, #3 (#2)}
\begin{abstract}

We study the resonant x-ray magnetic scattering (RXMS)
around the $K$ edge of Ni in the antiferromagnet NiO,
by treating the $4p$ states of Ni as a band and the $3d$ states 
as localized states.
We propose a mechanism that the $4p$ states are coupled 
to the magnetic order through the intra-atomic Coulomb interaction 
between the $4p$ and the $3d$ states
and through the $p$-$d$ mixing to the $3d$ states
of neighboring Ni atoms. 
These couplings induce the orbital moment in the $4p$ band,
and thereby give rise to the RXMS intensity at the $K$ edge
in the dipolar process. 
It is found that the spin-orbit interaction in the $4p$ band has
negligibly small contribution to the RXMS intensity.
The present model reproduces well the experimental spectra.
We also discuss the azimuthal angle dependence of the intensity.

\end{abstract}


\newpage

\section{introduction}

Resonant x-ray scattering has recently attracted much interest
as a useful tool to investigate magnetic and orbital orders.
The resonant enhancement for magnetic Bragg 
reflections has been observed in transition-metal 
compounds\cite{Hill,Neubeck1,Stunault,Neubeck2}
using synchrotron radiation with photon energy around the $K$ edge.
Also the resonant enhancement for the prohibited Bragg reflection
corresponding to the orbital order has been observed in LaMnO$_3$.
\cite{Murakami}
The resonant scattering is described by a second order process 
that a photon is virtually absorbed by exciting a core electron 
to unoccupied states 
and then emitted by recombining the excited electron with the core hole.
Therefore, for the $K$ edge, the $4p$ states are involved in the dipolar 
process, while the $3d$ states are involved in the quadrupolar process.
Since the $4p$ states are highly extended in space, 
they are rather sensitive to electronic structure at neighboring sites. 
This character is now well recognized in the resonant scattering
for the orbital order in LaMnO$_3$;
at the beginning, proposed was a mechanism that the intensity 
comes from
the $4p$ states modified by the intra-atomic Coulomb interaction 
with the $3d$ electrons constituting the orbital order,\cite{Ishihara} 
but the subsequent studies of treating the $4p$ states 
as a band\cite{Elfimov,Benfatto,Takahashi1,Takahashi2} have revealed that
the $4p$ states of Mn is mainly modified 
by the Jahn-Teller distortion via the oxygen potential on the neighboring 
sites, not by the intra-atomic Coulomb interaction.

In this paper, we study the mechanism of the resonant x-ray scattering 
for the magnetic order (RXMS) in NiO.
We analyze the recent experiment of Neubeck {\it et al.},\cite{Neubeck2} 
paying special attentions to the band effects of the $4p$ states.
In NiO, Ni atoms form an fcc lattice.
It is a type II antiferromagnet (AF) 
with the N\'eel temperature $T_{\rm N}=523$ K. 
The order parameter is characterized by a wavevector
${\bf Q}$, which is directed to one of four body-diagonals 
in the fcc lattice. Therefore there exist four domains 
($K$ domains).
The spin vector lies in the plane perpendicular 
to the AF modulation direction so that there are three possible directions 
($S$ domains).
For example, the spins are directed to $(-1,-1,2)$, $(2,-1,-1)$, 
or $(-1,2,-1)$ in the $K$ domain of 
${\bf Q}=(\frac{1}{2},\frac{1}{2},\frac{1}{2})$.
In total, we have 12 domains. 
In the RXMS experiment for NiO, the $K$ domain is specified,
but the $S$ domains are mixed.\cite{Neubeck2}

For generating the RXMS intensity, orbitals have to  coupled to 
the magnetic order; in the dipolar process, the $4p$ states have 
to couple to the $3d$ states.
Since such couplings are usually much smaller than those to 
the Jahn-Teller distortion in LaMnO$_3$,
the mechanism is much subtler than that for the orbital order in LaMnO$_3$.
The spin polarization alone in the $4p$ band
cannot give rise to the RXMS intensity; if the amplitudes of exciting electron 
are summed for up and down spins, the total becomes the same at each site 
in the absence of the orbital moment. This leads to the cancellation of
the amplitudes between different magnetic sublattices.
The spin-orbit interaction in the $4p$ band may break this cancellation,
\cite{Stunault,Neubeck2}
but its effect will be shown to be negligibly small.

As discussed in our previous paper for CoO,\cite{Igarashi1}
we propose that two couplings are effective to make the $4p$ states couple 
to the magnetic order: the intra-atomic Coulomb interaction between the $4p$
and $3d$ states, and the mixing of the $4p$ states
with the $3d$ states of neighboring Ni atoms.
To substantiate this observation, we use a tight-binding model 
for the $4p$ states, treating the $3d$ states within a cluster model.
We will show that these couplings actually give rise to the RXMS spectra
in agreement with the recent experiment.\cite{Neubeck2}
A caution is to be made on the intra-atomic Coulomb effect;
without treating the $4p$ states as a band,
the Coulomb interaction gives rise to too large RXMS intensities.
Thus the $4p$ band effect has a role of reducing the effect of
the Coulomb interaction.
We have also a clear peak in the pre-$K$-edge region,
which will be shown to come from the quadrupolar process. 
Since the $3d$ states strongly
couple to the magnetic order, having this peak is not surprising.
The present cluster model analysis, a slight extension of an atomic
model analysis,\cite{Neubeck2} reproduces well the experimental spectra.

This paper is organized as follows.
In Sec.~II, the present model is described, 
and the electronic structure is calculated on the basis of this model. 
In Sec.~III, the formulation of the RXMS spectra 
is developed, and the calculated results are discussed
in comparison with experiments. Section V is devoted to concluding
remarks. In the Appendix, the geometrical factors are given in
explicit forms.

\section{3D STATES} 

For describing the $3d$ states mixing to the $2p$ states of neighboring
O sites, we start by the following model Hamiltonian:
\begin{eqnarray}
H_0&=&\sum_{im\s}E_d(m)d_{im\s}^\dagger d_{im\s} 
     + H^I_{3d-3d} \nonumber\\
   &+&\zeta_{3d}\sum_{i}\sum_{mm'\s\s'}
      \langle m\s|{\bf\l}\cdot{\bf s}|m'\s'\rangle
      d_{im\s}^\dagger d_{im'\s'} \nonumber \\
   &+&\sum_{j\l\s}E_{2p}p_{j\l\s}^\dagger p_{j\l\s} 
     +\sum_{<i,j>}\sum_{\s\l m}
      (t^{2p,3d}_{im,j\l}p_{j\l\s}^dagger d_{im\s}+H.c.).
\end{eqnarray}
Operators $d_{im\s}$ and $p_{j\l\s}$ represent the annihilation of 
an electron with spin $\s$ in the $3d$ state $m$ of the Ni 
site $i$ and of an electron with spin $\s$
in the $2p$ state $\l$  of the O site $j$, respectively.
The first term describes the energy of the $3d$ level.
We take account of a point charge crystal-field splitting ($10Dq$) 
such that $E_d(e_g)=E_d+6Dq$, $E_d(t_{2g})=E_d-4Dq$.
The second term, $H^I_{3d-3d}$, represents the intra-atomic Coulomb interaction 
between the $3d$ states on Ni sites. Although the explicit form is
omitted here, the matrix elements are written in terms of the Slater integrals 
$F^0(3d,3d)$, $F^2(3d,3d)$, and $F^4(3d,3d)$.
The third term represents the spin-orbit interaction of the $3d$ states.
The fourth term represents the energy of the $2p$ level of O.
The relative energy is given by the charge-transfer energy $\Delta$ defined by 
$\Delta=E_d-E_{2p}+nU$ for the $d^n$ configuration with
$U=F^0(3d,3d)-\frac{2}{63}F^2(3d,3d)-\frac{2}{63}F^4(3d,3d)$.
The last term represents the mixing of the Ni $3d$ states with the O $2p$ 
states.  The mixing parameters $t_{im,j\l}^{2p,3d}$ are defined by 
the Slater-Koster integrals,
$(pd\s)_{2p,3d}$, $(pd\pi)_{2p,3d}$.\cite{Slater}
Most parameter values are assumed to be consistent with a previous 
cluster-model analysis\cite{Elp} and our previous electron correlation
calculation.\cite{Takahashi3} They are listed 
in Table \ref{tbl.param}. 
We evaluate the atomic values within the Hartree-Fock (HF) 
approximation.\cite{Cowan} 
For $F^0(3d,3d)$, we use a rather smaller value than the atomic one 
by considering screening effects in solids, 
while for $F^2(3d,3d)$ and $F^4(3d,3d)$ we use the atomic values
with multiplying a factor 0.8.

The ground state is calculated within this Hamiltonian as follows.
We assume that each cluster consisting of a Ni atom and surrounding six O
atoms is subjected to a molecular field coming from neighboring clusters;
the exchange energy in the mean-field approximation is given by
$H_{\rm ex}= J({\bf Q})\langle{\bf S}\rangle\cdot{\bf S}$,
where ${\bf S}$ represent the spin operator for the $3d$ states,
and $J({\bf Q})$ is the Fourier transform of the exchange constant
with the AF modulation vector ${\bf Q}$. 
Since the N\'eel temperature $T_{\rm N}$ is given by 
$T_{\rm N}=|J({\bf Q})|S(S+1)/(3k_{\rm B})$ in the mean field approximation,
the exchange energy is expressed as
\begin{equation}
 H_{\rm ex} = \frac{3|\langle{\bf S}\rangle|}{2S(S+1)}k_{\rm B} T_{\rm N}S_\eta,
\end{equation}
with the $\eta$ axis pointing to one of the directions $(-1,-1,2)$, 
$(2,-1,1)$, or $(-1,2,-1)$.
We put $S=1$, $|\langle{\bf S}\rangle|=1$ for Ni atoms,
and use the experimental value for $T_{\rm N}$.
Then we make numerical calculations, assuming that the ground state 
is given by a linear combination of the $3d^8$
configurations (45 states) and $3d^9\underline{L}$ configuration 
(360 states) in each cluster, where $\underline{L}$ indicates a state that
a hole is in the $2p$ states of neighboring O sites. 
Diagonalizing the Hamiltonian matrix ($405\times 405$ dimensions), 
we obtain the ground state that the weight of being in the $3d^8$ 
configuration is 0.865 with
the orbital moment $\langle L_\zeta\rangle=0.316$.
This value is consistent with the value $\langle L_\zeta\rangle=0.32$
in the recent polarization analysis of nonresonant magnetic-scattering 
\cite{Fernandez} and in the LDA$+U$ calculation.\cite{Kwon}
Note that the calculated value of the orbital moment is sensitive 
to the assumed value of $10Dq$.

In the quadrupolar process of the RXMS,
an electron is excited from the $1s$ state to the $3d$ states 
in the intermediate states.
Such excited states are numerically calculated in a similar way 
to that for the ground state.
We diagonalize the Hamiltonian matrix within the space of the $1s^1 3d^9$ 
configuration (10 states) and the $1s^1 3d^{10}\underline{L}$ configuration 
(36 states), assuming that an attractive interaction $V^c_{3d}\sim 4$ eV
works between the $3d$ electron and the core hole only 
in the core hole site.
The energy of the core-hole is adjusted such that a peak in the pre-K-edge 
region is located at $\hbar\omega=8330$ eV in the RXMS spectra (see Sec.~IV).

\section{4P STATES}

The RXMS involves a virtual process of a transition from the $1s$ state
to the $4p$ states at Ni sites in the dipolar process,
so that the $4p$ states are to be studied in detail.
This section is devoted to the study of the $4p$ states in connection with
the $K$-edge absorption spectra.

We use a tight-binding model for the $4p$ states.
Although many other states such as $4s$ state at Ni sites and $3s$ state
at O sites are mixed in such a rather-high-energy region, 
this model makes sense as an effective model for simply taking
account of the $4p$ band effects; the $4p$ states keep their atomic character
inside the atomic sphere even in solids.
As relevant couplings to the magnetic order, we introduce the intra-atomic 
Coulomb interaction between the $4p$ and the $3d$ states, 
and the mixing of the $4p$ states with the $3d$ states in neighboring Ni atoms. 
Thereby a part of the Hamiltonian concerning the $4p$ states is given by
\begin{eqnarray}
H_1&=&\sum_{in\s}E_{4p}a_{in\s}^\dagger a_{in\s} 
    +\sum_{<i,i'>}\sum_{\s nn'}\left(
      t'^{4p,4p}_{in,i'n'}a^\dagger_{in\s}a_{i'n'\s} +{\rm H. c.}\right)
\nonumber\\
   &+&\zeta_{4p}\sum_{i}\sum_{nn'\s\s'}
      \langle n\s|{\bf\l}\cdot{\bf s}|n'\s'\rangle
      a_{in\s}^\dagger a_{in'\s'} \nonumber\\
   &+& H^I_{4p-3d} 
   +\sum_{\s nm}\left(t'^{4p,3d}_{in,i'm}a^\dagger_{in\s}d_{i'm\s}+H.c.\right),
\label{eq.4p}
\end{eqnarray}
where $a_{in\s}$ represents the annihilation of
an electron with spin $\s$ in the $4p$ state $n$ on Ni site $i$.
The first and the second terms describe the $4p$ bands.
The transfer integral $t'^{4p,4p}_{in,i'n'}$ is written 
in terms of the Slater-Koster parameters, $(pp\s)_{4p,4p}$ and 
$(pp\pi)_{4p,4p}$.\cite{Slater} We have the $4p$ band width $\sim 20$ eV,
assuming $(pp\pi)_{4p,4p}\sim 3.5$ eV for the nearest-neighbor Ni-Ni distance,
and the half of this value for the next-nearest-neighbor distance, with
an empirical relation $(pp\pi)_{4p,4p}\sim -(pp\sigma)_{4p,4p}/4$.
The value of $E_{4p}$ is adjusted such that the peak in the $K$-edge
absorption is located at $\hbar\omega=8437$ eV.
The third term represents the spin-orbit interaction in the $4p$ states.  
The HF atomic value is used for $\zeta_{4p}$. 
The fourth term $H^I_{4p-3d}$ represents the intra-atomic Coulomb interaction
between the $4p$ states and the $3d$ states. 
The matrix elements are written in terms of the Slater integrals, 
$F^0(4p,3d)$, $F^2(4p,3d)$, $G^1(4p,3d)$, and $G^3(4p,3d)$. 
By the same reason as in $H^I_{3d-3d}$, the HF atomic values are used
for  $F^2(4p,3d)$, $G^1(4p,3d)$, and $G^3(4p,3d)$ with multiplying 
a factor 0.8. The value of $F^0(4p,3d)$ is rather ambiguous, since it can be
absorbed into the value of $E_{4p}$.
The last term describes the mixing of the $4p$ states with the $3d$ states
of neighboring sites. The mixing parameter $t'^{4p,3d}_{in,i'm}$ is written 
in terms of the Slater-Koster parameters, $(pd\s)_{4p,3d}$ and 
$(pd\pi)_{4p,3d}$. We assume $(pd\s)_{4p,3d}\sim -1.0$ eV together with
an empirical relation $(pd\pi)_{4p,3d}\sim -(pd\sigma)_{4p,3d}/2$.
Although this mixing is much smaller than the transfer between the $4p$
states, it is not neglected in the RXMS spectra, since it makes
the $4p$ states couple to the magnetic order.
The parameter values are listed in Table \ref{tbl.param}.

We numerically calculate the $4p$ band as follows.
The $H_1$ together with $H_0$ is represented in each Ni site
within the space of states that one electron is in the $4p$ states 
with the $3d$ electrons in the ground state (6 states) 
and that a $3d$ electron is added to become the $3d^9$-configuration 
(10 states).
Such states in different Ni sites are coupled through the second and
the last terms in Eq.~(\ref{eq.4p}), so that they form a band.
Considering two Ni atoms in the unit cell due to the AF order,
the Hamiltonian matrix has $32\times 32$ dimensions for each wave vector,
which is numerically diagonalized.

In the final state of the $K$-edge absorption and the intermediate states
of RXMS, there is a hole in the $1s$ state, which attracts the $4p$ electron.
Assuming that the core-hole potential works only within the core-hole site,
we solve the potential problem with respect to a local Green's function
defined by
\begin{equation}
 \Pi_{0n\s;0n\s}(\omega) = -i\int_{-\infty}^{\infty}
             \langle T\left(b_{0\s}^\dagger(t)a_{0n\s}(t)
             a_{0n'\s'}^\dagger b_{0\s'}(0)\right)\rangle 
             {\rm e}^{i\omega t}{\rm d}t,
\end{equation}
where $b_{0\s}$ is the annihilation operator of the $1s$ state with spin $\s$
at site $0$, $\langle\cdots\rangle$ the average over the ground state, 
and $T$ the time ordering.
Taking account of the multiple scattering due to the core-hole potential,
we have
\begin{equation}
\Pi_{0n\s;0n\s}(\omega)
  = \sum_{n''\s''}\Pi_{0n\s;0n''\s''}^{(0)}(\omega)
    [I -V^c_{4p}\Pi^{(0)}(\omega)]^{-1}_{0n''\s'';0n'\s'},
\end{equation}
with
\begin{equation}
 \Pi_{0n\s;0n\s}^{(0)}(\omega)
  = \sum_R\frac{\langle 0n\s|R\rangle
                          \langle R| 0n'\s'\rangle}
           {\omega - (E_R+\epsilon_{\rm core}-E_g)+i\Gamma/2} .
\end{equation}
Here the state $|R\rangle$ represents the $4p$ band with energy $E_R$
discussed above, and $\epsilon_{\rm core}$ is the core hole energy.
The life-time broadening width $\Gamma$ and the core-hole potential 
$V^c_{4p}$ are set to be $1.7$ eV,\cite{Neubeck2} and $3$ eV, respectively.

The spectral function of the local Green's function,
$I(\omega)=\sum_{n\s}(-1/\pi){\rm Im}\,\Pi_{0n\s;0n\s}(\omega)$,
is proportional to the $K$-edge absorption spectra and the fluorescence 
spectra.
Figure \ref{fig.absorp} shows $I(\omega)$
in comparison with the $K$ edge fluorescence experiment.\cite{Neubeck2}
The solid and the broken lines represent the spectra for $V^c_{4p}=0$ and
$3$ eV, respectively.
The attractive interaction enhances the intensity in the low-energy side.
It consists of two peaks. The peak in low-energy region corresponds to
the $K$-edge peak in the experimental spectra.
We are mainly interested in the energy region of this peak
in the RXMS spectra.
Th peak in high energy region may correspond to the peak around
$\hbar\omega=8365$ eV in the experimental spectra, but the calculated
peak is too large.
This discrepancy may come from using the $4p$ tight-binding model;
in reality, it is likely to become much broader through the mixing 
to other states such as the $4s$ state of Ni and $3s$ states of O.

\section{RESONANT X-RAY MAGENTIC SCATTERING}

\subsection{Procedure of calculation}

The magnetic scattering amplitude has been reviewed by several authors.
\cite{deBergevin1,Blume1,Blume2}
We summarize here the expression of the amplitude in a form 
suitable to the present calculation.
The scattering geometry is shown in Fig.~\ref{fig.geom};
photon with frequency $\omega$, momentum ${\bf k}_i$ and polarization $\mu$ 
($=\sigma$ or $\pi$) is scattered into the state
with momentum ${\bf k}_f$ and polarization $\mu'$ ($=\sigma'$ or $\pi'$). 
For the system with two magnetic sublattices $A$ and $B$ in the AF order,
the cross section for elastic scattering at magnetic Bragg peaks is given by
\begin{equation}
 \left. \frac{d\sigma}{d\Omega'}\right|_{\mu\to\mu'} \propto
  \left| T^A_{\mu\to\mu'}({\bf G},\omega) 
       - T^B_{\mu\to\mu'}({\bf G},\omega) \right|^2 ,
\end{equation}
with
\begin{eqnarray}
  T^{\eta}_{\mu\to\mu'}({\bf G},\omega) =
    &-&\frac{i\hbar\omega}{mc^2}\left(\frac{1}{2}{\bf L}^{\eta}({\bf G})
      \cdot {\bf A}''+{\bf S}^{\eta}({\bf G})\cdot{\bf B}\right)\nonumber\\
    &+&J^{\eta}_{\mu\to\mu'}({\bf G},\omega)
     +L^{\eta}_{\mu\to\mu'}({\bf G},\omega).
\label{eq.amp}
\end{eqnarray}
where ${\bf G}={\bf k}_f-{\bf k}_i$ is the scattering vector.

The first term in eq.~(\ref{eq.amp}) represents a {\em non-resonant} term,
where $m$ and $c$ are the electron mass and the velocity of photon, 
respectively.\cite{Blume2}
${\bf A}''$ and ${\bf B}$ are given by
\begin{eqnarray}
 {\bf A}''&=& {\bf A}'-({\bf A}'\cdot{\bf\hat G}){\bf\hat G},
  \quad {\bf A}'=-4\sin^2\theta(\hat{\mbox{\boldmath $\epsilon$}}'\times
  \hat{\mbox{\boldmath $\epsilon$}}),\\
 {\bf B} &=& \hat{\mbox{\boldmath $\epsilon$}}'\times 
 \hat{\mbox{\boldmath $\epsilon$}}
             + ({\bf\hat k}_f\times \hat{\mbox{\boldmath $\epsilon$}}')
               ({\bf\hat k}_f\cdot \hat{\mbox{\boldmath $\epsilon$}})
             - ({\bf\hat k}_i\times \hat{\mbox{\boldmath $\epsilon$}})
               ({\bf\hat k}_i\cdot \hat{\mbox{\boldmath $\epsilon$}}')
             - ({\bf\hat k}_f\times \hat{\mbox{\boldmath $\epsilon$}}')
        \times ({\bf\hat k}_i\times \hat{\mbox{\boldmath $\epsilon$}}),
\end{eqnarray}
where $\hat{\mbox{\boldmath $\epsilon$}}$ and 
$\hat{\mbox{\boldmath $\epsilon$}}'$ are the initial
and scattered polarizations, and
${\bf\hat k}_i$, ${\bf\hat k}_f$, and ${\bf\hat G}$ are normalized 
vectors of ${\bf k}_i$, ${\bf k}_f$, and ${\bf G}$.
${\bf L}^\eta ({\bf G})$ and ${\bf S}^\eta({\bf G})$ stand for the Fourier
transform of the orbital and spin angular momentum densities,
which values are known from the recent non-resonant RXS experiment:
\cite{Neubeck2,Fernandez}
$|{\bf L}({\bf G})| (|{\bf S}({\bf G})|) = 0.26\,(0.82)$, $0.24\,(0.60)$,
and $0.12\,(0.20)$ for ${\bf G}=(\frac{1}{2},\frac{1}{2},\frac{1}{2})$,
$(\frac{3}{2},\frac{3}{2},\frac{3}{2})$, and 
$(\frac{5}{2},\frac{5}{2},\frac{5}{2})$, respectively.
We insert these experimental values into Eq.~(\ref{eq.amp}).
The local moments of orbital and spin are give by
${\bf L}^\eta ({\bf G})$ and ${\bf S}^\eta({\bf G})$ with ${\bf G}\to 0$.
The above experimental values are consistent with the present 
ground-state calculation in Sec.~II.

The second term in Eq.~(\ref{eq.amp}) represents
the resonant contribution from the dipolar process. 
It is expressed by a second-order process that a photon is virtually absorbed 
by exciting the $1s$ electron to the $4p$ states and then emitted
by recombining the excited electron with the $1s$ core hole.
The corresponding amplitude is expressed as 
\begin{equation}
  J_{\mu\to\mu'}^\eta({\bf G},\omega)
  = \sum_{\alpha\alpha'}(P'^{\mu'})_{\alpha}
    M_{\alpha\alpha'}^\eta(\omega)P^{\mu}_{\alpha'},
\label{eq.damp}
\end{equation}
with the scattering tensor 
\begin{equation}
 M_{\alpha\alpha'}^\eta(\omega) = \sum_{\Lambda} 
  \frac{m\omega_{\Lambda g}^2\langle g|x_\alpha|\Lambda\rangle
  \langle \Lambda|x_{\alpha'}|g\rangle}
       {\hbar\omega-(E_{\Lambda}-E_g)+i\Gamma/2}.
\label{eq.dipole}
\end{equation}
The geometrical factors $P^\mu$, $P'^{\mu'}$ are given in explicit
forms in the Appendix. In Eq.~(\ref{eq.dipole}),
$\omega_{\Lambda g}=(E_{\Lambda}-E_g)/\hbar$, and
$x_\alpha$'s stand for the dipole operators, $x_1=x$, $x_2=y$,
and $x_3=z$ in the coordinate frame fixed to the crystal axes. 
The ground state $|g\rangle$ has an energy $E_g$.
The intermediate state $|\Lambda\rangle$ consists of an excited electron 
on the $4p$ states and a hole on the $1s$ state with energy $E_{\Lambda}$.
In the presence of the core-hole potential, it is convenient to
rewrite the scattering tensor in terms of the local Green's function
defined in Sec.~III. We have
\begin{equation}
 M^\eta_{\alpha\alpha'}(\omega)= A_p^2m\omega_c^2
  \sum_{\s}\Pi_{0\alpha\s,0\alpha\s}(\omega) ,
\end{equation}
where site $0$ is to belong to sublattice $\eta$, and 
$\hbar\omega_c\approx 8350$ eV. The $A_p$ is the dipole matrix element 
defined by $A_p=\langle 4p|r|1s\rangle =
   \int_0^{\infty} R_{4p}(r)rR_{1s}(r)r^2{\rm d}r $
with $R_{1s}(r)$ and $R_{4p}(r)$ being the radial wavefunctions 
for the $1s$ and $4p$ states.
We evaluate it as $A_p=1.52\times 10^{-11}$ cm
within the HF approximation\cite{Cowan} in the $3d^8$ configuration of 
a free Ni ion.
The amplitudes at A and B sublattices are nearly cancelled out,
leading to an antisymmetric form for the difference:
\begin{equation}
 M^A(\omega)-M^B(\omega)
   = \left( \begin{array}{rrr}
            0 & a & c \\
           -a & 0 & b \\
           -c &-b & 0
            \end{array} \right) ,
\end{equation}
with $a$, $b$, and $c$ being some complex numbers.

The third term in Eq.~(\ref{eq.amp}) represents
the resonant contribution from the quadrupolar process. 
The corresponding amplitude is expressed as 
\begin{equation}
 L_{\mu\to\mu'}^\eta({\bf G},\omega)
  = \sum_{\gamma\gamma'}(Q'^{\mu'})_{\gamma}
    N_{\gamma\gamma'}^\eta(\omega)Q^{\mu}_{\gamma'},
\label{eq.qamp}
\end{equation}
with the scattering tensor,
\begin{equation}
 N_{\gamma\gamma'}^\eta(\omega) = \frac{k^2}{12}\sum_{\Lambda'}
  \frac{m\omega_{\Lambda'g}^2\langle g|z_\gamma|\Lambda'\rangle\langle \Lambda'
       |z_{\gamma'}|g\rangle}
       {\hbar\omega-(E_{\Lambda'}-E_g)+i\Gamma/2}.
\label{eq.quadrupole}
\end{equation}
The geometrical factors, $Q^\mu$, $Q'^{\mu'}$, are given in explicit 
forms in the Appendix.
In Eq.~(\ref{eq.quadrupole}), $k$ is the wavevector of the incident 
(and scattered) photon, $\sim 4.2\times 10^8$ cm$^{-1}$,
$\omega_{\Lambda' g}=(E_{\Lambda'} - E_g)/\hbar$,
and $z_\mu$'s stand for the quadrupole operators,
$z_1=(\sqrt{3}/2)(x^2-y^2)$, $z_2=(1/2)(3z^2-r^2)$, $z_3=\sqrt{3}yz$,
$z_4=\sqrt{3}zx$, and $z_5=\sqrt{3}xy$ in the coordinate frame
fixed to the crystal axes.
The intermediate states $|\Lambda'\rangle$ consist of an excited electron 
on the $3d$ states and a hole on the $1s$ states with
energy $E_{\Lambda'}$, which are evaluated by following the procedure 
described in Sec~II.
The scattering amplitude $N^\eta_{\gamma\gamma'}(\omega)$ contains the
square of the quadrupole matrix element 
$A_d = \langle 3d|r^2|1s\rangle =
   \int_0^{\infty} R_{3d}(r)r^2R_{1s}(r)r^2{\rm d}r$,
with $R_{3d}(r)$ being the radial wavefunction of a free Ni ion.
We evaluate it as $A_d = 2.10\times 10^{-20}$ cm$^2$ 
within the HF approximation in the $3d^8$ configuration.
Similar to the dipolar process,
the scattering tensors $N^A(\omega)$ and $N^B(\omega)$ are nearly 
cancelled out, leading to an antisymmetric form for the difference:
\begin{equation}
 N^A(\omega)-N^B(\omega)
   = \left( \begin{array}{rrrrr}
             0 & d & e & f & g \\
            -d & 0 & h & p & q \\
            -e &-h & 0 & r & s \\
            -f &-p &-r & 0 & t \\ 
            -g &-q &-s &-t & 0 
          \end{array} \right) , 
\end{equation} 
with $d\sim t$ being some complex numbers.  
Note that this difference would vanish if the spin-orbit interaction
in the $3d$ states were absent. This is consistent with the prediction 
by Loversey.
\cite{Loversey}

\subsection{Calculated results}

In this subsection, we discuss the RXMS spectra 
in comparison with the recent experiment of Neubeck {\it et al.}\cite{Neubeck2}
In their experiment, the $K$ domain of 
${\bf Q}=(\frac{1}{2},\frac{1}{2},\frac{1}{2})$
is selected, but the contributions from different $S$ domains are not 
separated. Fortunately the relative volumes for three $S$ domains
have been determined such that $0.55\,(S_1)$, $0.4\,(S_2)$, $0.05\,(S_3)$ for
${\bf Q}=(\frac{1}{2},\frac{1}{2},\frac{1}{2})$,
$0.63 (S_1)$, $0.37 (S_2)$, $0 (S_3)$ for
${\bf Q}=(\frac{3}{2},\frac{3}{2},\frac{3}{2})$,
and $0.26 (S_1)$, $0.33 (S_2)$, $0.41 (S_3)$ for
${\bf Q}=(\frac{5}{2},\frac{5}{2},\frac{5}{2})$.
In the following, we average the spectra over three $S$ domains
multiplying the weights of these relative volumes. 

Figure \ref{fig.spec1} shows the spectra thus calculated
as a function of photon energy
for ${\bf G}=(\frac{1}{2},\frac{1}{2},\frac{1}{2})$.
The azimuthal angle $\psi$ is set to be zero such that 
the scattering plane contains the $(1,-1,0)$ crystal axis, 
corresponding to the experimental situation.
The left panels show the spectra for the $\sigma\to\sigma'$ channel.
Only one peak is found at $\hbar\omega=8330$ eV in the pre-$K$-edge region, 
in agreement with the experiment. 
Since the dipolar process is forbidden in this polarization, 
this peak arises from the quadrupolar process.
A strong anti-resonant dip is seen in the low energy side of the peak,
which comes from an interference of the non-resonant process
with the quadrupolar process.
The experimental intensity in the high energy side of the peak
is much smaller than the calculated intensity.
This discrepancy may be related to the lack of absorption correction
on the experimental spectra,
which correction should make the intensity in the high energy side the same
as that in the low energy side.
This observation conversely suggests the size of the absorption correction.
The right panels show the spectra for the $\sigma\to\pi'$ channel.
Similar to the $\sigma\to\sigma'$ channel, we find a sharp peak
at $\hbar\omega=8330$ eV, which arises from the quadrupolar process. 
The intensity for the energy far below $\hbar\omega=8330$ eV
is much smaller than the value for the $\sigma\to\sigma'$ channel.
In addition, we have several structures in the high-energy side of the peak.
As plotted by the long-dashed line, the non-resonant process and 
the quadrupolar process cannot give rise to sufficient intensities, 
indicating that the intensity mainly comes from the dipolar process.
We have checked the effect of the spin-orbit interaction in the $4p$ band
by calculating the spectra with neglecting that interaction;
no visible change is found.
We have a peak around $\hbar\omega = 8346$ eV, in agreement with the experiment.
To this peak, the intra-atomic Coulomb interaction between the $4p$ 
and the $3d$ states contributes substantially (compare the solid line
with the dashed one). 
The experimental intensity, which is smaller than the calculated one,
should be enhanced by absorption correction. 
We should emphasize here that 
the dipolar contribution becomes more than $10$ times as large as 
the present values, too large in comparison with the experimental one, 
if the transfer of the $4p$ electron to neighboring sites is neglected.
\cite{Com2}
Therefore, the $4p$ band effect is very important for
reproducing the experimental spectra.
A large peak at $\hbar\omega\sim 8365$ eV is regarded as
an artifact of using the tight-binding model, as mentioned in Sec.~III.

Figure \ref{fig.spec2} shows the spectra as a function of photon energy,
at $\psi=0$ for ${\bf G}=(\frac{3}{2},\frac{3}{2},\frac{3}{2})$.
The left panels show the spectra for the $\sigma\to\sigma'$ channel.
The spectral shape is quite similar to that for 
${\bf G}=(\frac{1}{2},\frac{1}{2},\frac{1}{2})$.
The right panels show the spectra for the $\sigma\to\pi'$ channel.
The non-resonant intensity is larger than that for
${\bf G}=(\frac{1}{2},\frac{1}{2},\frac{1}{2})$.
We find a sharp peak at $\hbar\omega = 8330$ eV, which arises from
the quadrupolar process.
In the high-energy side of this peak, we also have several structures
arising from the dipolar process;
a peak around $\hbar\omega = 8346$ eV agrees well with the experiment.

Figure \ref{fig.spec3} shows the spectra as a function of photon energy 
for ${\bf G}=(\frac{5}{2},\frac{5}{2},\frac{5}{2})$.
The azimuthal angle $\psi$ is now set to be $270^\circ$, corresponding to
the experimental situation. 
The right panels show the spectra for the $\sigma\to\pi'$ channel.
For the peak at $\hbar\omega\sim 8330$ eV,
an anti-resonant dip now moved to the high-energy side. 
We also find a broad peak around $\hbar\omega\sim 8340$ eV, owing to
the dipolar process.

Figure \ref{fig.azim} shows the spectra as a function of the azimuthal 
angle $\psi$ for ${\bf G}=(\frac{1}{2},\frac{1}{2},\frac{1}{2})$,
in comparison with the experiment.
The lowest panel shows the intensity of the peak at $\hbar\omega=8346$ eV 
in the main-$K$-edge region for the $\sigma\to\pi'$ channel. 
We multiply a factor such that the calculated values at $\psi=0$ 
becomes close to the experimental one.
The variation resembles to a dependence of $\sin^2\psi$,
consistent with the dominant contribution of the dipolar process.
The top and middle panels show the intensities of the peaks
at $\hbar\omega=8330$ eV in the pre-$K$-edge region for $\sigma\to\sigma'$ 
and the $\sigma\to\pi'$ channels, respectively.
We multiply a same factor in both cases such that the values at $\psi=0$
become close to the experimental ones.
The variation in the top panel looks simply a $90^\circ$-shift 
of the curve in the lowest panel. 
This is a little surprising, since we expect a more complex shape
from the quadrupolar process.
The curve in the middle panel has an extra peak. This may be related 
to the fact that the dipolar contribution remains in this peak.
Such an extra peak is not clear in the experiment.
\section{CONCLUDING REMARKS}

We have studied the RXMS around the $K$ edge in the antiferromagnet NiO,
treating the $4p$ states as a band and the $3d$ states as localized
states. We have proposed a mechanism that the $4p$ states are coupled
to the magnetic order through the intra-atomic Coulomb interaction
between the $4p$ and $3d$ states, and through
the $p$-$d$ mixing to the $3d$ states of neighboring Ni atoms.
Thereby the orbital moment is induced in the $4p$ states,
giving rise to the RXMS intensity in the dipolar process.
The calculated spectra in the present model have reproduced well
the experimental ones.\cite{Neubeck2}
Different from previous analyses,\cite{Stunault,Neubeck2}
the effect of the spin-orbit interaction in the $4p$ band is found
negligibly small within the present model.
This reminds us of the magnetic circular dichroism (MCD)
for the $K$-edge absorption in the ferromagnetic metals Fe, Co, and Ni.
\cite{Schutz} 
It is known that the MCD intensity is mainly generated by
the orbital moment induced in the $4p$ band 
through the mixing to the $3d$ states of neighboring atoms, 
not through the spin-orbit interaction in the $4p$ band.\cite{Igarashi2}

We have also obtained a peak in the pre-$K$-edge region,
in agreement with a previous analysis in the atomic model.\cite{Neubeck2} 
This arises from the quadrupole process, showing a contrast to 
the pre-$K$-edge peak in the resonant scattering 
for the orbital order in LaMnO$_3$. The latter is predicted to be generated 
by the dipolar process,\cite{Takahashi2}
where the states of $p$ symmetry (seen from the Mn atoms) have weights
in the pre-$K$-edge region by mixing to the $3d$ states of neighboring
Mn atoms. Usually the intensity of the quadrupolar process is much smaller 
than that of the dipolar process, 
but the dipolar contribution cannot become large in the present case
because of a weak coupling of the $4p$ states to the magnetic order.

Furthermore, we have discussed the azimuthal-angle dependence of the spectra.
The curve for the pre-$K$-edge peak, which comes from the quadrupolar 
process, looks simply a $90^\circ$-shift of the curve for 
the main-$K$-edge peak, which comes mainly from the dipolar process.

Finally, we would like to stress again that the tight-binding model is a 
first approach for taking account of the $4p$ band effects. 
More realistic band calculations are desirable for obtaining more 
quantitative results. 

\acknowledgments

We would like to thank W. Neubeck for kindly informing
the experimental data for NiO prior to publication. 
This work was partially supported by 
a Grant-in-Aid for Scientific Research 
from the Ministry of Education, Science, Sports and Culture, Japan.

\appendix
\section{Geometrical factor}

We summarize here the expressions of geometrical factors for the
dipolar and quadrupolar processes. The detailed derivation is given in
Ref.~[11].
For the scattering vectors ${\bf G}=(\frac{1}{2},\frac{1}{2},\frac{1}{2})$, 
$(\frac{3}{2},\frac{3}{2},\frac{3}{2})$, and 
$(\frac{5}{2},\frac{5}{2},\frac{5}{2})$,
the geometrical factor $P^\mu$ ($\mu=\sigma$ or $\pi$) 
for incident photon in the dipolar process is given by 

\begin{eqnarray}
 \left(P^\sigma\right)_1 &=& (\cos\beta\cos\psi + \sin\psi)/\sqrt{2},
                              \nonumber\\
 \left(P^\sigma\right)_2 &=& (\cos\beta\cos\psi - \sin\psi)/\sqrt{2},
                              \nonumber\\
 \left(P^\sigma\right)_3 &=& -\sin\beta\cos\psi,
\label{eq.dsigma} \\
 \left(P^\pi\right)_1 &=& \left[\sin\theta(\cos\beta\sin\psi-\cos\psi)
                                +\cos\theta\sin\beta\right]/\sqrt{2},
                                \nonumber\\
 \left(P^\pi\right)_2 &=& \left[\sin\theta(\cos\beta\sin\psi+\cos\psi)
                                +\cos\theta\sin\beta\right]/\sqrt{2},
                                \nonumber\\
 \left(P^\pi\right)_3 &=& (-\sin\theta\sin\beta\sin\psi
                                +\cos\theta\cos\beta),
\label{eq.dpi}
\end{eqnarray}
where with $\beta=\arccos\sqrt{1/3}$, and $\theta$ represents 
the Bragg angle. The azimuthal angle
$\psi=0$ corresponds to the scattering plane containing 
the $(1,-1,0)$ crystal axis. 
The geometrical factor $P'^{\mu'}$ ($\mu'=\sigma'$ or $\pi'$) for 
the scattered photon is given by replacing $\sin\theta$ by $-\sin\theta$
in Eqs.~(\ref{eq.dsigma}) and (\ref{eq.dpi}).

Also, the geometrical factor  $Q^\mu$ ($\mu=\sigma$ or $\pi$)
for incident photon in the quadrupolar process is given by
\begin{eqnarray}
 \left(Q^\sigma\right)_1 &=& 
    -\cos\theta\cos\beta\cos 2\psi
    -\sin\theta\sin\beta\sin\psi,
    \nonumber\\
 \left(Q^\sigma\right)_2 &=& 
    (\sqrt{3}/2)\cos\theta\sin^2\beta\sin 2\psi
    +(\sqrt{3}/2)\sin\theta\sin 2\beta\cos\psi,
    \nonumber\\
 \left(Q^\sigma\right)_3 &=& 
    -\cos\theta\sin\beta(\cos\beta\sin 2\psi+\cos 2\psi)/\sqrt{2}\nonumber\\
   &-&\sin\theta(\cos 2\beta\cos\psi-\cos\beta\sin\psi)/\sqrt{2},
    \nonumber\\
 \left(Q^\sigma\right)_4 &=& 
    -\cos\theta\sin\beta(\cos\beta\sin 2\psi-\cos 2\psi)/\sqrt{2}\nonumber\\
   &-&\sin\theta(\cos 2\beta\cos\psi+\cos\beta\sin\psi)/\sqrt{2},
    \nonumber\\
 \left(Q^\sigma\right)_5 &=& 
    \cos\theta[1-(1/2)\sin^2\beta]\sin 2\psi
    -(1/2)\sin\theta\sin 2\beta\cos\psi,
\label{eq.qsigma} \\
 \left(Q^\pi\right)_1 &=& 
    -(1/2)\sin 2\theta\cos\beta\sin 2\psi
    -\cos 2\theta\sin\beta\cos\psi,
    \nonumber\\
 \left(Q^\pi\right)_2 &=& 
    -(\sqrt{3}/4)\sin 2\theta\sin^2\beta\cos 2\psi
    -(\sqrt{3}/2)\cos 2\theta\sin 2\beta\sin\psi\nonumber\\
   &-&(\sqrt{3}/2)\sin 2\theta[1-(3/2)\sin^2\beta],
    \nonumber\\
 \left(Q^\pi\right)_3 &=& 
    (1/2\sqrt{2})\sin 2\theta\sin\beta(\cos\beta\cos 2\psi-\sin 2\psi)
    \nonumber\\
   &+&\cos 2\theta(\cos 2\beta\sin\psi+\cos\beta\cos\psi)/\sqrt{2}\nonumber\\
   &-&(3/4\sqrt{2})\sin 2\theta\sin 2\beta,
    \nonumber\\
 \left(Q^\pi\right)_4 &=& 
    (1/2\sqrt{2})\sin 2\theta\sin\beta(\cos\beta\cos 2\psi+\sin 2\psi)
    \nonumber\\
   &+&\cos 2\theta(\cos 2\beta\sin\psi-\cos\beta\cos\psi)/\sqrt{2}\nonumber\\
   &-&(3/4\sqrt{2})\sin 2\theta\sin 2\beta,
    \nonumber\\
 \left(Q^\pi\right)_5 &=& 
    -(1/2)\sin 2\theta[1-(1/2)\sin^2\beta]\cos 2\psi
    \nonumber\\
   &+&(1/2)\cos 2\theta\sin 2\beta\sin\psi
    -(3/4)\sin 2\theta\sin^2\beta.
\label{eq.qpi}
\end{eqnarray}
The geometrical factor $Q'^{\mu'}$ ($\mu'=\sigma'$ or $\pi'$) 
for scattered photon is given by replacing $\sin\theta$ 
and $\sin 2\theta$ by $-\sin\theta$ and $-\sin 2\theta$, respectively,
in Eqs.~(\ref{eq.qsigma}) and (\ref{eq.qpi}).

\begin{table}
\caption{Parameter values for the tight-binding model of NiO 
in units of eV.\label{tbl.param}}
\begin{tabular}{ccccc}
 Parameter$^a$       &         & Parameter$^b$      &    \\ \hline
 $F^0_{dd}$          &$ 6.5   $&$F^0_{pd}$          &$ 2.5  $\\ 
 $F^2_{dd}$          &$ 9.8   $&$F^2_{pd}$          &$ 1.9  $\\
 $F^4_{dd}$          &$ 6.1   $&$G^1_{pd}$          &$ 0.68 $\\
 $\Delta$            &$ 6.0   $&$G^3_{pd}$          &$ 0.59 $\\
 $(pd\sigma)_{2p,3d}$&$-1.4   $&$(pp\sigma)_{4p,4p}$&$ 3.5  $\\
 $(pd\pi)_{2p,3d}$   &$ 0.63  $&$(pp\pi)_{4p,4p}$   &$-0.88 $\\
 $10Dq$              &$ 0.50  $&$(pd\sigma)_{4p,3d}$&$-1.0  $\\ 
                     &         &$(pd\pi)_{4p,3d}$   &$ 0.50 $\\ 
 $\zeta_{3d}$        &$0.096\,(0.12^c$)&$\zeta_{4p}$&$0.098$\\ 
\end{tabular}
\leftline{$^a$ Values in the $1s^2 3d^8$ configuration.}
\leftline{$^b$ Values in the $1s^1 3d^8 4p^1$ configuration.}
\leftline{$^a$ Value in the $1s^1 3d^9$ configuration.}
\end{table}

\begin{figure}

\caption{
Finite section of a NiO crystal. Only Ni atoms are shown.
Wave vector ${\bf Q}=(\frac{1}{2},\frac{1}{2},\frac{1}{2})$
characterizes an antiferromagnetic modulation direction.
Three arrows in a plane perpendicular to ${\bf Q}$ indicate
possible spin directions ($S$ domains).
Spins for solid, grey, and open circles align alternately. 
\label{fig.cryst}}

\vskip 20pt

\caption{
Spectral function $I(\omega)$ for the local Green's function for the $4p$
states (lower panel), in comparison with the $K$-edge fluorescence spectra
(upper panel).$^{4}$
Solid and broken lines represent the spectra calculated with 
the core-hole potential $V^c_{4p}=0$ and $3$ eV, respectively.
\label{fig.absorp}}

\vskip 20pt

\caption{
Scattering geometry of x ray scattering. Incident photon with
wave vector ${\bf k}_i$ and polarization $\sigma$ or $\pi$ 
is scattered into the state with wave vector ${\bf k}_f$ and
polarization $\sigma'$ or $\pi'$ at Bragg angle $\theta$.
The sample crystal is rotated by azimuthal angle $\psi$
around the scattering vector ${\bf G}={\bf k}_f-{\bf k}_i$.
\label{fig.geom}}

\vskip 20pt

\caption{
Magnetic scattering spectra for 
${\bf G}=(\frac{1}{2},\frac{1}{2},\frac{1}{2})$
at $\psi=0^\circ$, as a function of photon energy,
in comparison with the experiment.
The upper panels show the experimental spectra without absorption
correction.$^{4}$
The left and right panels show the intensity for the $\sigma\to\sigma'$
channel, and that for the $\sigma\to\pi'$ channel, respectively.
In the right and lower panel, the dashed line represents the intensity
by neglecting the intra-atomic Coulomb interaction between the $4p$ and
the $3d$ states, and the long-dashed line represents the contribution
from the non-resonant and the quadrupolar processes.
\label{fig.spec1}}

\vskip 20pt

\caption{
Magnetic scattering spectra for 
${\bf G}=(\frac{3}{2},\frac{3}{2},\frac{3}{2})$
at $\psi=0^\circ$, as a function of photon energy,
in comparison with the experiment.
The upper panels show the experimental spectra without absorption
correction.$^{4}$
The left and right panels show the intensity for the $\sigma\to\sigma'$
channel and that for the $\sigma\to\pi'$ channel, respectively.
\label{fig.spec2}}

\vskip 20pt

\caption{
Magnetic scattering spectra for 
${\bf G}=(\frac{5}{2},\frac{5}{2},\frac{5}{2})$
at $\psi=270^\circ$, as a function of photon energy,
in comparison with the experiment.
The upper panels show the experimental spectra without absorption
correction.$^{4}$
The left and right panels show the intensity for the $\sigma\to\sigma'$
channel and that for the $\sigma\to\pi'$ channel, respectively.
\label{fig.spec3}}

\vskip 20pt

\caption{
Magnetic scattering intensity as a function of azimuthal angle
$\psi$, for ${\bf G}=(\frac{1}{2},\frac{1}{2},\frac{1}{2})$,
in comparison with the experiment. The open circles represent
the experimental intensities.$^{4}$
The top and middle panels are for the peak at $\hbar\omega=8330$ eV
in the pre-$K$-edge region for the $\sigma\to\sigma'$ and 
$\sigma\to\pi'$ channels, respectively.
The lowest panel is for the peak at $\hbar\omega=8346$ eV 
in the main-$K$-edge region for the $\sigma\to\pi'$ channel.
\label{fig.azim}}

\end{figure}


\begin{references}
\bibitem{Com2} 
In the previous analysis of the RXMS spectra for CoO within a cluster model
(Ref.~[11]), we had to assume that the values of the intra-atomic Coulomb 
interaction is considerably smaller than the atomic values for
obtaining the reasonable (not too large) RXMS intensities.
\end{references}
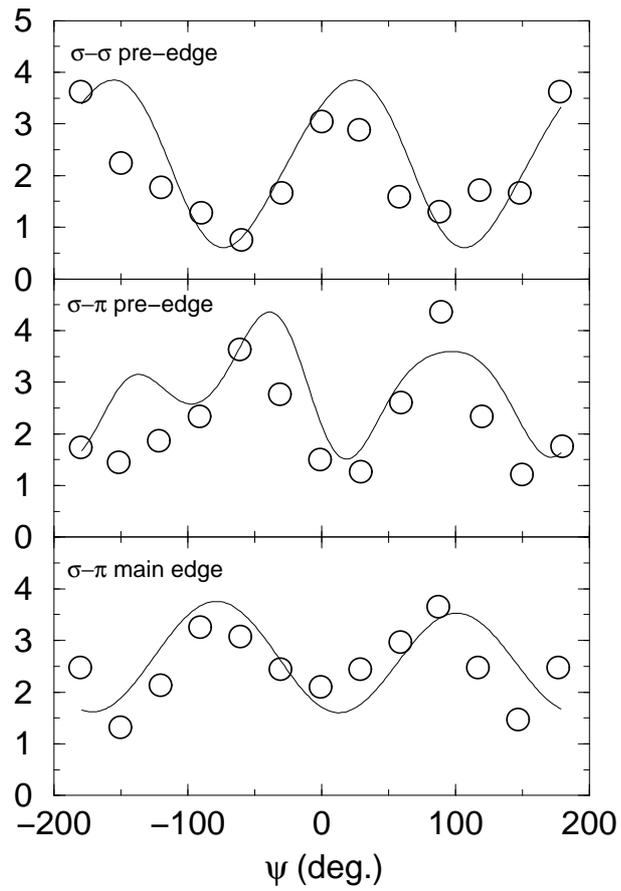
\end{document}